\documentclass[aip,rsi,reprint]{revtex4-2}
\usepackage{amsmath}
\usepackage{amsfonts}
\usepackage{amssymb}
\usepackage{graphicx}
\usepackage{color}
\usepackage{tikz}
\usepackage{pgfplots}
\usetikzlibrary{external}
\pgfplotsset{compat=1.3}

\begin{document}


\title{Air-photonics terahertz platform with versatile micro-controller based interface and data acquisition} 



\author{E. Prost}
\email{emilien.prost@univ-lyon1.fr}
\affiliation{Univ Lyon, Université Claude Bernard Lyon 1, CNRS, Institut Lumière Matière, F-69622, Villeurbanne, France}
\author{V. Loriot}
\affiliation{Univ Lyon, Université Claude Bernard Lyon 1, CNRS, Institut Lumière Matière, F-69622, Villeurbanne, France}
\author{E. Constant}
\affiliation{Univ Lyon, Université Claude Bernard Lyon 1, CNRS, Institut Lumière Matière, F-69622, Villeurbanne, France}
\author{I. Compagnon}
\affiliation{Univ Lyon, Université Claude Bernard Lyon 1, CNRS, Institut Lumière Matière, F-69622, Villeurbanne, France}
\author{L. Berg\'e}
\affiliation{CEA, DAM, DIF, F-91297 Arpajon, France}
\affiliation{Universit\'e Paris-Saclay, CEA, LMCE, 91680 Bruy\`eres-le-Ch\^atel, France}
\author{F. Lépine}
\affiliation{Univ Lyon, Université Claude Bernard Lyon 1, CNRS, Institut Lumière Matière, F-69622, Villeurbanne, France}
\author{S. Skupin}
\affiliation{Univ Lyon, Université Claude Bernard Lyon 1, CNRS, Institut Lumière Matière, F-69622, Villeurbanne, France}


\date{\today}

\begin{abstract}
We present a recently developed terahertz platform.
An air plasma produced by an ultrashort two-color laser pulse serves as a broadband terahertz source, with electric field that has peak amplitude in the MV/cm range.
Air biased coherent detection of the terahertz field is employed where a peak detector associated with a micro-controller board acquires the signal coming from an avalanche photodiode.
The temporal and spectral profiles of the produced terahertz electric field are presented, in excellent agreement with numerical simulations of the whole setup.
We illustrate the capabilities of this platform by performing spectroscopy on water vapor and of a polystyrene reference sample.
\end{abstract}


\maketitle 

\section{Introduction}
The terahertz (THz) part of the electromagnetic spectrum, usually defined between 0.1~THz and 30~THz, has attracted great interest in various research fields in the last decades.
This recent progress in THz science has led to the advent of numerous applications~\cite{Tonouchi2007,RedoSanchez2013}.
Terahertz time domain spectroscopy~\cite{Jepsen2011,Xie2014,Neu2018} (THz-TDS) is one of those and it is extensively used in different domains of research ranging from material science~\cite{Hangyo2005} to chemistry~\cite{Fischer_2005, BAWUAH2021116272}.
Nowadays, commercial THz-TDS systems are available, making this technique a widespread tool for numerous applications (see~\cite{Neu2018} and references therein).
One of the usual limits of THz-TDS systems is the bandwidth of the THz emitters and detectors as well as the amplitude of the produced fields.
Usually, emitters and detectors rely on photoconductive antennas or optical rectification in non-linear crystals with bandwidths limited to 10~THz and suffering from bandwidth gaps due to absorption by optical phonons in the material.

The improvement of THz-TDS systems goes hand in hand with the development of appropriate THz sources~\cite{Hafez2016,Zhang2021}.
An interesting alternative source relies on the use of an air plasma generated by a two color laser pulse~\cite{Cook2000, Kress2004, Kim2007}. Such THz source can be conveniently coupled with air biased coherent detection~\cite{Karpowicz2008,Ho2010} (ABCD).
Using air as the generation and detection medium has several advantages, most notably the ability to produce broadband high amplitude THz fields without any risk of damaging the source or detector.
Moreover, the propagation and detection of these broadband THz spectra was demonstrated, with bandwidths going up to 75~THz\cite{Kim2008} and beyond\cite{doi:10.1063/1.4732524}.
Not relying on solid material also means avoiding the appearance of gaps in the THz spectra due to optical phonons.

Here we report on the development of a THz platform based on the above mentioned air plasma source and ABCD setup.
In contrast to previous works using lock-in amplifiers in the detection process, our data acquisition scheme relies on the use of a peak detector to collect the data and a micro-controller board as an analog to digital converter.
The first part of this paper is dedicated to the presentation of the layout of the platform.
After describing the optical setup and the signal detection, we introduce the micro-controller based data acquisition and interface of the platform. 
In the second part of the paper we present experimental results measured with the platform.
We begin by showing the temporal and spectral envelope of the produced THz pulses.
Numerical simulations of the platform including the THz source and the detection support our measurements and reveal important differences between generated and measured THz spectra.
Then we demonstrate the spectroscopic capabilities of the platform.
First by measuring the fine structure imprinted by water vapor on the spectrum.
Secondly, by using a polystyrene reference sample and comparing the results to measurements obtained with a Fourier transform infrared (FTIR) spectrometer.

\section{THz platform setup}
\subsection{Optical setup}
Our optical setup relies on THz generation by a two-color ultra-short laser pulse\cite{Cook2000, Kress2004, Kim2007} composed of the fundamental harmonic (FH) and its second harmonic (SH).
The key process driving the THz generation has been identified as an asymmetric current caused by photo-ionization and subsequent acceleration of free electrons ~\cite{Kim2008,Babushkin2010}.
The produced THz field is measured with an ABCD setup\cite{Dai2006, Karpowicz2008, Lu2009} where the THz pulse and a synchronized FH laser pulse are focused and superposed in the presence of a DC bias field.
Four wave mixing due to the $\chi^{(3)}$ non-linearity of air produces a SH radiation, a process also called terahertz field induced second harmonic generation\cite{Cook1999}(TFISH).
In order to be able to retrieve the THz electric field from the SH pulse, an additional high voltage DC bias field is applied during the four wave mixing process.
The produced SH signal and its treatment are discussed hereafter.

\begin{figure}[!tbp]
\tikzsetnextfilename{setup}
\input{figures/setup.tikz}
\caption{Scheme of the experimental setup. BS : Beam splitter (50\%); L : focusing lens ($f=300$~mm); DL : Delay line; HWP : Half waveplate; BH : composed of a BBO crystal and an half-waveplate; BD : Beam dump; SiW : Silicon wafer; S : Sample; HV : High voltage electrodes; F : 400~nm band-pass filter; APD : Avalanche photodiode. The gray area highlights the part of the setup within an air controlled enclosure.}
\label{Fig:Setup}
\end{figure}

The optical setup is presented in Fig.~\ref{Fig:Setup}.
It uses 20\% of the output of a Ti:sapphire chirp-pulse amplifier (Coherent Legend Elite Duo HP USX) producing 30~fs pulses at a repetition rate of 5~kHz with 2~mJ per pulse at a central wavelength of 800~nm.
The 2~W incoming beam is split equally between two arms, the first one being dedicated to THz generation and the second one to THz detection.
The full THz beam-path is placed in a dry air enclosure (see gray area in Fig.~\ref{Fig:Setup}), in order to prevent the strong absorption bands of water in the THz range.

The SH component of our two-color laser pulses is produced in a 100~µm thick $\beta$-Barium Borate (BBO) crystal adjusted for type I phase matching and a subsequent ultra-thin half-waveplate at 800~nm (WPD03-H800-F400-SP from Newlight Photonics).
The second harmonic generation (SHG) process converts part of the fundamental pulse at the frequency $\omega$ into its SH at frequency $2\omega$.
Optimized type I SHG produces a SH field with its polarization axis orthogonal to the FH field.
The THz yield is much larger when both components have parallel polarization\cite{Kosareva:18}, therefore we use a half-waveplate to align both the FH and SH polarization axis.
The use of an ultra-thin waveplate is mandatory to keep some temporal overlap between the two colors. 
The BBO crystal and the waveplate are placed 70~mm before the geometrical focus of a 300~mm lens.
Another half-waveplate is placed before the lens to adjust the polarization axis of the FH field such that the polarization axis of the produced linearly polarized THz field and the FH probe pulse are parallel in the detection area.

A few millimeters before the geometrical focus of the lens a plasma is generated by the two color laser pulse and serves as an elongated THz point source.
The emitted THz field and the two color laser pulse are collected and collimated by an off axis parabolic mirror.
After this first mirror the two color optical pulse is removed by a silicon wafer that transmits about 50\% of the THz power\cite{Kaltenecker2019}.
The transmitted THz field is focused using a second off axis parabolic mirror, before being collimated by a third one.
The samples are placed at that intermediate THz focus.
A motorized sample holder allows us to perform measurements of several samples without opening the dry air enclosure.

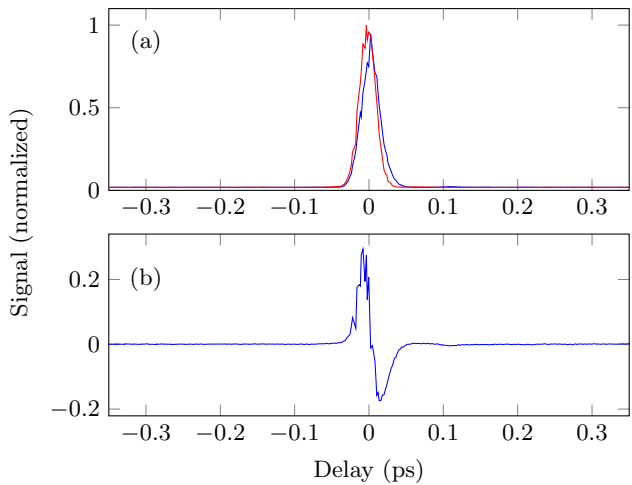
\begin{figure}[!htbp]
\tikzsetnextfilename{Heterodynedetection}
\begin{tikzpicture}
\begin{axis}[
height = 4cm,
width = 8.5cm,
legend style={draw=none},
xmin = -0.35,
xmax = 0.35,
ymin = 0]
\addplot[color=blue] table [x = delay, y expr= \thisrow{sig_neg}, mark = none] {figures/Data_heterodynedetection.txt};
\addplot[color=red] table [x = delay, y expr= \thisrow{sig_pos}, mark = none] {figures/Data_heterodynedetection.txt};
\node[] at (axis cs: -0.3,0.9) {(a)};
\legend{$S_{2\omega}^{-}$,$S_{2\omega}^{+}$};
\end{axis}
\begin{axis}[
height = 4cm,
width = 8.5cm,
legend style={draw=none},
xlabel={Delay (ps)},
ylabel = {Signal (normalized)},
ylabel style={xshift=1.5cm, yshift = 0cm},
xmin = -0.35,
xmax = 0.35,
yshift = -3cm]
\addplot[color = black] table [x = delay, y expr=\thisrow{sig_het}, mark = none] {figures/Data_heterodynedetection.txt};
\node[] at (axis cs: -0.3,0.2) {(b)};
\legend{$S_{2\omega}^{+}-S_{2\omega}^{-}$};
\end{axis}
\end{tikzpicture}
\caption{(a) Second harmonic signal measured by applying a bias field $E_{\mathrm{bias}}$ of  $10$~kV/cm with positive (red) or negative (blue) sign. (b) Reconstructed terahertz field amplitude from the difference between the two previous measurements. All signals are normalized to the maximum measured signal.}
\label{Fig:HeterodyneDetection}
\end{figure}

The probe part of the FH laser field propagating through a delay line (DL325 from Newport) is dedicated to THz detection.
The THz pulse is focused again using the last off axis parabolic mirror.
A hole is this mirror allows the FH probe pulse, at frequency $\omega$, to be focused by a 300~mm lens at the same position as the THz pulse.
At focus, a static 2~kV bias voltage is applied along the probe and THz polarization direction by a pair of 2~mm long electrodes separated by 2~mm.
This results in TFISH through a four wave mixing process between the probe FH field $E_{\omega}$ and the low frequency electric field $E_{\Omega} = E_{\mathrm{THz}}+E_{\mathrm{bias}}$ where $E_{\mathrm{THz}}$ and $E_{\mathrm{bias}}$ are the THz and bias fields, respectively.
We use a bandpass filter (FB400-40 from Thorlabs) centered at 400~nm to isolate the produced SH signal, which is then detected by an avalanche photodiode (APD440A2 from Thorlabs).
The probe and low frequency fields are both time dependent and the resulting SH signal depends on their cross-correlation.
The SH signal $S_{2\omega}$ received by the photodiode, for a given time delay $\tau$ between the probe and THz field, can be expressed as\cite{Karpowicz2008}
\begin{equation*}
S_{2\omega}(\tau) \propto  \left( \chi^{(3)}I_{\omega} \right)^{2} \left[ E_{\mathrm{THz}}^{2}(\tau) + E_{\mathrm{bias}}^{2} + 2E_{\mathrm{bias}}E_{\mathrm{THz}}(\tau)\right],
\end{equation*}
where $\chi^{(3)}$ denotes the effective third-order susceptibility of air and $I_{\omega}$ the intensity of the probe pulse.
Usually, THz-TDS relies on the use of an external low-frequency modulation of $E_{\mathrm{bias}}$ and lock-in amplification to extract the THz electric field from the SH signal.
Here, we choose a different approach that requires to perform two successive measurements,  $S_{2\omega}^+$ and $S_{2\omega}^-$, with bias field of opposite sign, $E_{\mathrm{bias}}=\pm 10$~kV/cm, as shown in Fig.~\ref{Fig:HeterodyneDetection}(a).
By subtracting the two signals $S_{2\omega}^+$ and $S_{2\omega}^-$ we obtain a signal $S_{\mathrm{THz}}$ directly proportional to the THz electric field,
\begin{equation*}
S_{\mathrm{THz}}(\tau) = S_{2\omega}^+(\tau) -  S_{2\omega}^-(\tau) \propto  4\left( \chi^{(3)}I_{\omega} \right)^{2} \left|E_{\mathrm{bias}}\right|E_{\mathrm{THz}}(\tau).
\end{equation*}
Note that by comparing the magnitude of SH signals with and without bias field, the THz field amplitude can be estimated when $E_{\mathrm{bias}}$ is known\cite{Iwaszczuk2012}.

\subsection{Acquisition and interface electronics}

\begin{figure}[!tbp]
\tikzsetnextfilename{Detection_scheme}
\begin{tikzpicture}
\fill[gray!20] (-0.8,1.5) rectangle (0.8,2.5);
\node[align = center,text width=1.6cm] at (0,2) {Computer};

\fill[blue!20] (1.7,1.5) rectangle (3.3,2.5);
\node[align = center,text width=1.6cm] at (2.5,2) {Delay line};

\fill[gray!20] (0,0) circle (0.8);
\node[align = center,text width=1.6cm] at (0,0) {Arduino};

\fill[red!20] (1.5,0) -- (2.5,0.8) -- (3.5,0) -- (2.5,-0.8) -- cycle;
\node[align = center,text width=1.6cm] at (2.55,0) {Peak detector};

\fill[blue!20] (-1.8,-2.6) rectangle (-3.4,-1.4);
\node[align = center,text width=1.6cm] at (-2.6,-2) {High voltage supply};

\fill[blue!20] (-0.05,-2.6) rectangle (-1.7,-1.4);
\node[align = center,text width=1.6cm] at (-0.85,-2) {Motorized beamblock};

\fill[blue!20] (0.05,-2.6) rectangle (1.7,-1.4);
\node[align = center,text width=1.6cm] at (0.85,-2) {Humidity sensor};

\fill[blue!20] (1.8,-2.6) rectangle (3.4,-1.4);
\node[align = center,text width=1.6cm] at (2.6,-2) {Avalanche photodiode};

\draw[<->,thick] (0,0.8) -- (0,1.5); 
\draw[<->,thick] (0.8,2) -- (1.7,2); 
\draw[<->,thick] (0.8,0) -- (1.5,0); 
\draw[->,thick] (-0.8,0) -- (-2.6,-1.4); 
\draw[->,thick] (-0.05,-0.8) -- (-0.85,-1.4); 
\draw[<-,thick] (0.05,-0.8) -- (0.85,-1.4); 
\draw[<-,thick] (2.5,-0.8) -- (2.6,-1.4); 
\end{tikzpicture}
\caption{Scheme of the interface and data acquisition. The arrows represent the direction of the data flow.}
\label{Fig:Detection_scheme}
\end{figure}
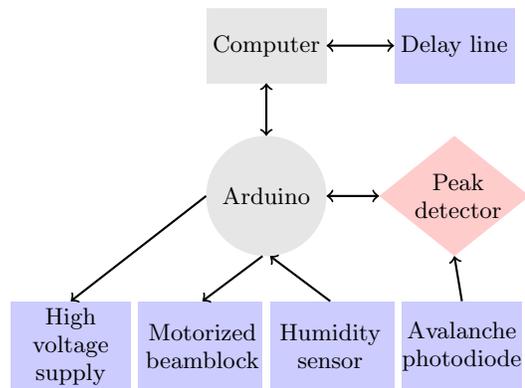

A schematic diagram of the interface and data acquisitions used in our THz platform is presented in Fig.~\ref{Fig:Detection_scheme}.
The core of the electronics is a micro-controller board (Arduino UNO Rev3) that serves as an interface between various elements of the platform and the computer.
We use the LabVIEW Hobbyist Toolkit library\cite{makerhub} to control and interact with the micro-controller.
The only other element directly controlled by the computer is the delay line, all other  communications are performed via the micro-controller.
The desired sign of the applied bias field is sent to the high voltage supply.
Motorized beamblocks, composed of a piece of anodized aluminum fixed on a servomotor (Parallax Standard Servo \#900-00005), are placed on both pump and probe arms.
Their positions are controlled by pulse-width modulation and can thus be controlled by the micro-controller.
The Arduino board does not only control all other elements but also serves as an analog to digital converter.
We use it to record the experiment signal as well as the humidity inside the enclosure with a sensor (Honeywell HIH-4000-001) placed near the sample holder.
The measured signal does not come directly from the photodiode but from a homemade peak detector\cite{Achtenberg2020}, which is used to convert the photodiode signal into a DC voltage readable by the micro-controller board.
At its output, the avalanche photodiode delivers a voltage pulse, whose amplitude is proportional to the power of the received light pulse, with a maximum value of 2V.
The peak detector takes the highest value of this input electronic signal and stores it as a constant output voltage.
Before being processed, this output voltage is amplified by a factor 2.5 to use the whole dynamic range of the analog to digital converter of the micro-controller.
A reset of the peak detector is performed before each measurement.

\section{THz platform results}

\subsection{THz pulses produced by the platform}
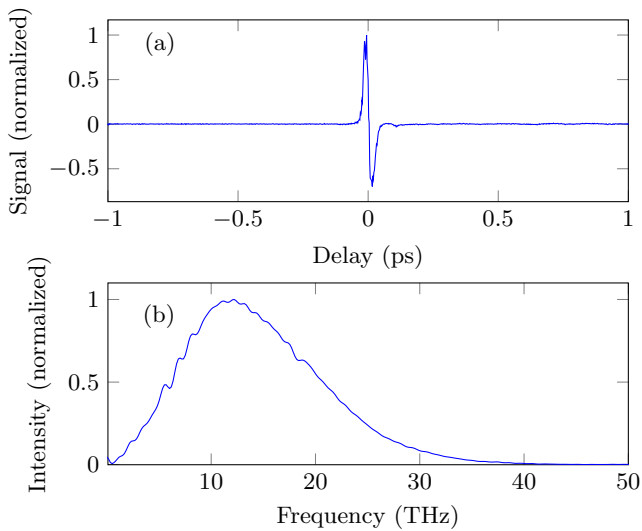
\begin{figure}[!tbp]
\tikzsetnextfilename{Profile}
\begin{tikzpicture}
\begin{axis}[
height = 4cm,
width = 8.5cm,
xlabel={Delay (ps)},
ylabel = {Signal (normalized)},
xmin = -1,
xmax = 1]
\addplot table [x = delay, y expr= \thisrow{sig}/0.73169, blue, mark = none] {figures/Data_profile_temp.txt};
\node[] at (axis cs: -0.8,0.9) {(a)};
\end{axis}
\begin{axis}[
height = 4cm,
width = 8.5cm,
xlabel={Frequency (THz)},
ylabel = {Intensity (normalized)},
xmin = 0.1,
xmax = 50,
ymin = 0,
yshift = -3.5cm]
\addplot table [x = nu, y expr= \thisrow{spec}/88.2524, mark = none] {figures/Data_profile_spec.txt};
\node[] at (axis cs: 5,0.9) {(b)};
\end{axis}
\end{tikzpicture}
\caption{Measured (a) temporal profile and (b) reconstructed spectral intensity of the THz electric field.}
\label{Fig:Profile}
\end{figure}

Figure~\ref{Fig:Profile} shows experimentally measured temporal and spectral profiles of the THz pulses produced by the platform.
For those measurements, no sample was placed in the path of the THz field and the relative humidity inside the enclosure was kept around 9\% by injecting dry air.
The measured temporal profile is short, with an envelope around 50~fs, and seems to be an almost single cycle pulse.
Using measurements taken with a known applied bias voltage of 10~kV/cm and without any bias\cite{Iwaszczuk2012}, the THz electric field strength has been evaluated in the MV/cm range. 
The THz spectrum is obtained by taking the Fourier transform of the measured temporal electric field.
Thus, the chosen temporal step and span for the acquisition also defines the spectral ones.
For the measurement presented in Fig.~\ref{Fig:Profile}, the temporal step is set to 2~fs and the data are taken over a 2~ps span giving a 0.5~THz spectral step and a 250~THz span.
These settings are well suited to measure the spectral envelope of the THz field and the shape of the temporal THz field.
The spectral intensity profile shown in Fig.~\ref{Fig:Profile}(b) has a smooth profile extending from 1~THz to 35~THz.
This measurement was performed by averaging over 10 laser shots for each data point leading to a total acquisition time of a couple of minutes.

\subsection{Numerical simulations}

\begin{figure*}[!tbp]
\includegraphics[width=\textwidth]{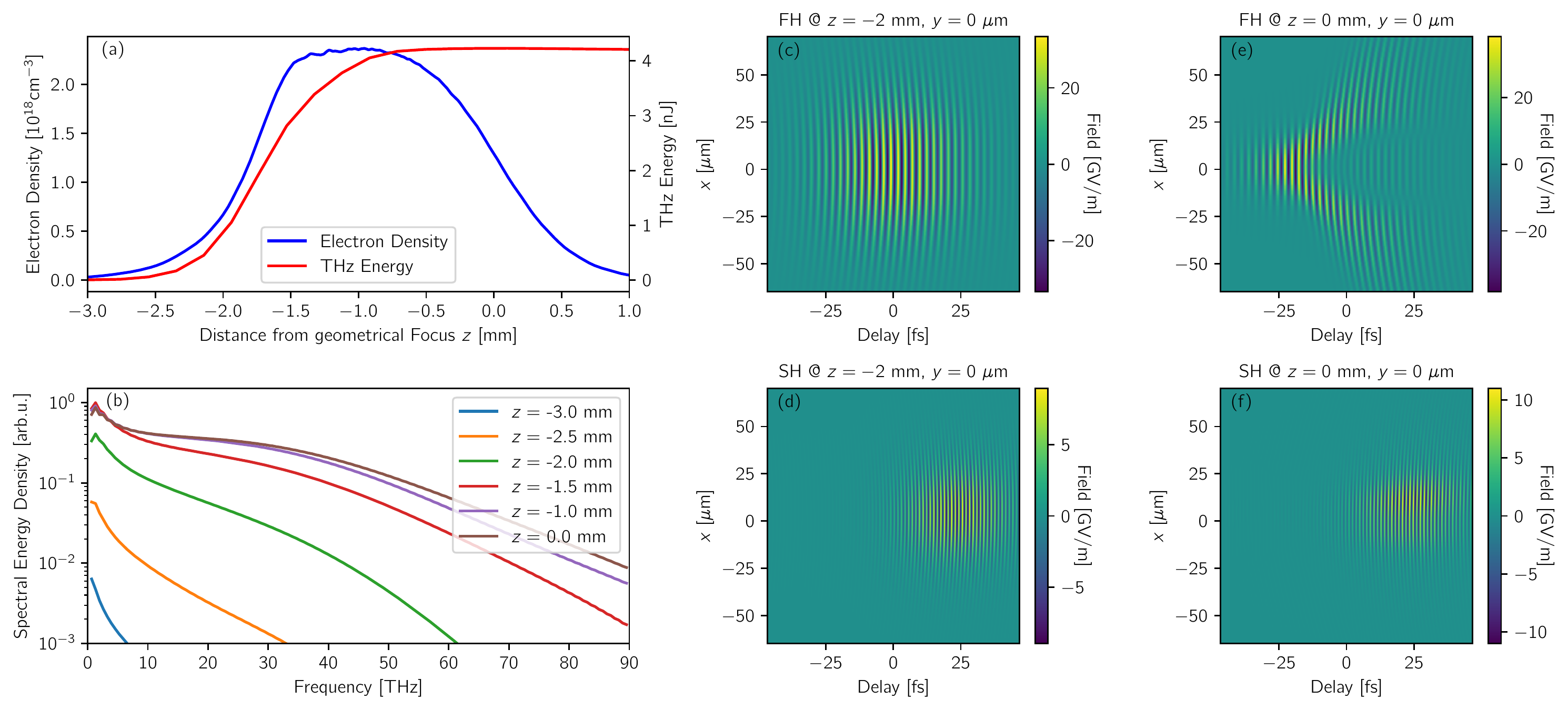}
\caption{THz source: (a) Peak plasma density and THz energy ($\nu<90$~THz) versus propagation distance relative to the geometrical focus at $z=0$. (b) Snapshots of the transverse integrated THz spectrum at several positions $z$ along the plasma. Snapshots of the main FH (c) and SH (d) polarization component of the pump field near the beginning of the plasma at $z=-2$~mm. (e) and (f) show the same fields near the end of the plasma at the geometrical focus $z=0$.}
\label{Fig:numerics}
\end{figure*}

To understand the different stages of the THz platform and the results presented in the previous section, we also performed numerical simulations in the exact same experimental conditions.
For the description of the air-plasma based THz source, we use direct fully space and time resolved computations based on a vectorial version of the unidirectional pulse propagation equation (UPPE)~\cite{Kolesik:pre:70:036604}.
This numerical model governs the forward-propagating transverse electric field components $E_x$, $E_y$ of elliptically-polarized pulses, which are subject to linear dispersion and diffraction; together with third-order nonlinear polarization, photo-ionization and related losses~\cite{Berge:rpp:70:1633}.
More detail about this vectorial model can be found in \cite {Tailliez:njp:22:103038} while the physical parameters for air are taken from \cite{Nguyen:oe:2017}.

In the simulations presented thereafter, we used our experimental laser parameters, namely an incident linearly polarized $115~\mu$J laser pulse at~800~nm with 30~fs FWHM duration and 6~mm FWHM beam width on the focusing lens L.
This fundamental harmonic (FH) pulse is then frequency doubled in the BBO crystal to create a co-propagating second harmonic (SH) pulse ($2.7\%$ energy conversion).
The polarizations of the FH and SH pulses are aligned by means of a half-waveplate before the pump intensity reaches values above the threshold for ionization and a two-color air-plasma is generated about 2~mm before the geometrical focus, here assumed at $z=0$.
The simulated peak electron density as a function of the propagation distance $z$ is shown in Fig.~\ref{Fig:numerics}(a), together with the energy build-up of the generated spectral components below 90~THz.
Figure~\ref{Fig:numerics}(b) reveals that the generated THz spectrum broadens significantly during the generation process, with higher frequencies generated later, that is, closer to the end of the plasma.

It turns out to be crucial to take into account the dispersion properties of the BBO, the quartz half-waveplate as well as the ambient air in order to get the correct beam and pulse profile of the SH component as well as its (slightly elliptic) polarization state at focus.
Figure~\ref{Fig:numerics}(c)-(f) shows the FH and SH field components at the beginning ($z=-2$~mm) and at the end ($z=0$) of the interaction region.
In particular the temporal walk-off between the two components is clearly visible, as the SH pulse is delayed by about 25~fs compared to the FH pulse.
While at $z=-2$~mm some overlap near zero delay still enables the photocurrent mechanism for THz generation, plasma defocusing in the trailing part of the FH pulse has completely separated the two colors in the spatio-temporal domain.
This is why THz generation stops roughly in the middle of the plasma at $z=-1$~mm, and the expected THz energy yield remains in the order of few nJ.

\begin{figure*}[!tbp]
\includegraphics[width=\textwidth]{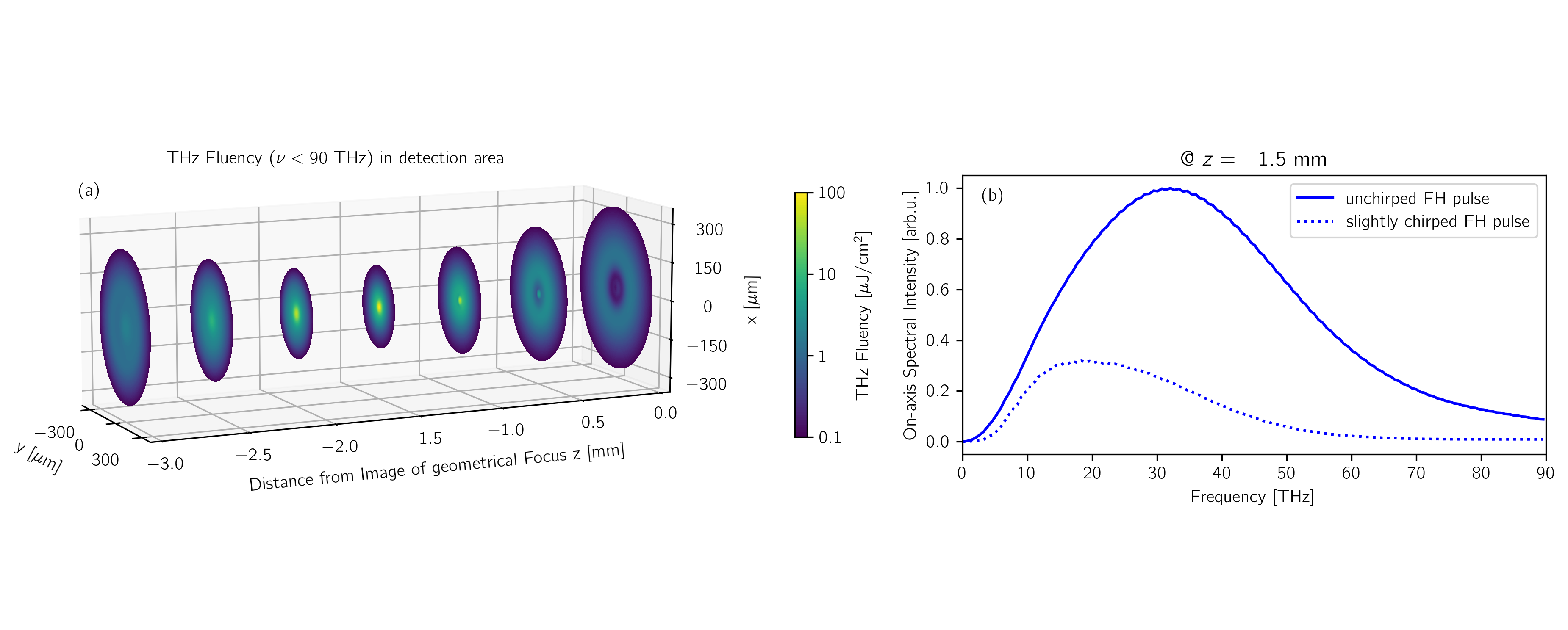}
\caption{THz detection: (a) THz fluency ($\nu<90$~THz) in the detection area between the HV electrodes as a function of the propagation distance. The position of the image of the geometrical focus of the pump laser is denoted with $z=0$. (b) Resulting on-axis ($x=y=0$) spectral intensity at the position of maximum THz fluency $z=-1.5$~mm in the detection area. For comparison, the same quantity is plotted for a slightly down-chirped input FH pulse with a group delay dispersion of $\mathrm{GDD}=-100$~fs$^2$.}
\label{Fig:numerics_detection}
\end{figure*}

Finally, we want to shed some light on the THz detection process, that is, the part of the generated THz field that we can expect to detect in our measurements.
One could incorrectly expect to measure a pattern close to the final spectrum at the end of the plasma (brown curve z=0~mm) in Fig.~\ref{Fig:numerics}(b).
This is not the case because those spectra are integrated over the whole transverse plane.
In the experimental detection setup only the part of the THz field overlapping with the focused IR probe pulse and the high voltage field is detected.
Therefore, it is instructive to compute the THz field in the detection area.
The four parabolic mirrors are essentially forming two consecutive 4f systems and thus an image of the THz source is produced.
Figure~\ref{Fig:numerics_detection}(a) shows the evolution of the THz fluence as a function of the propagation distance in the detection area (accounting for 50\% transmittance of the silicon wafer).
For convenience, the position of the image of the geometrical focus of the laser pump is set to $z=0$.
The highest THz fluency (and intensity) is obtained at $z=-1.5$~mm, which corresponds to the position of maximum THz generation of the source.
It is reasonable to focus the IR probe pulse at this position in order to maximize the ABCD signal.
Consistently with our measurements the IR focal FWHM spot size can be estimated to less than $20~\mu$m, that is, much smaller than the THz spot-size, and thus essentially the on-axis THz intensity is detected. 

The simulated on-axis THz spectral intensity at $z=-1.5$~mm in the detection area is shown in Fig.~\ref{Fig:numerics_detection}(b).
Compared to the spectrum emitted by the source shown in Fig.~\ref{Fig:numerics}(b) lower THz frequencies are strongly suppressed.
The reason for the suppression of lower THz frequencies in the ABCD scheme is that the THz spot size is strongly frequency dependent, as it is limited by the wavelength.
Thus, for those THz frequencies the on-axis intensity decreases.
Moreover, beam divergence increases with decreasing frequency, and parts of the low frequency field may not be collected by the parabolic mirror.

It is important to note that the air-plasma based THz source is very sensitive to the two-color pump waveform.
This is demonstrated by adding small chirp to the incoming FH pulse.
This chirp is imprinted to the SH pulse in the generation process in the BBO, and as a result both frequency and phase relation of FH and SH in the region of temporal overlap in the plasma is affected.
Even a small chirp corresponding to a group delay dispersion of $-100$~fs$^{2}$ will alter the recorded THz spectrum significantly, as shown by the dotted curve in Fig.~\ref{Fig:numerics_detection}(b).
Both amplitude and width of the spectrum are strongly decreased, rendering the simulated spectrum much closer to the experimental one presented in Fig.~\ref{Fig:Profile}(b).
A chirp of this order of magnitude is likely to be present in the upcoming experiments.
Despite this, our measurements are reproducible, thus showing the stability of the platform.

\subsection{Spectroscopic features of the THz pulses}
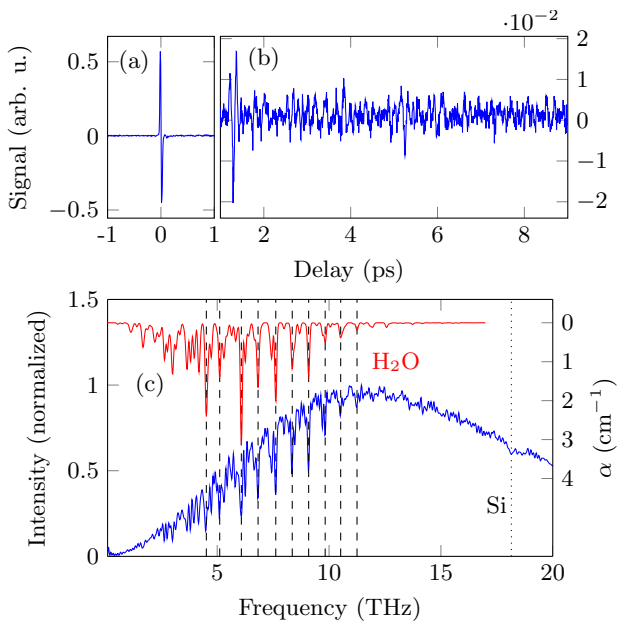
\begin{figure}[!tbp]
\tikzsetnextfilename{Spectrum}
\begin{tikzpicture}
\begin{axis}[
height = 4cm,
width = 3cm,
xlabel={Delay (ps)},
ylabel = {Signal (normalized)},
xlabel style={xshift=2.5cm, yshift = 0cm},
xmin = -1,
xmax = 1,
xtick ={-1,0,1},
xticklabels = {-1,0,1}]
\addplot table [x = delay, y expr= \thisrow{sig}/0.7093, mark = none] {figures/Data_spectrum_temp.txt};
\node[] at (axis cs: -0.5,0.9) {(a)};
\end{axis}
\begin{axis}[
height = 4cm,
width = 6.2cm,
/pgfplots/tick scale binop={\times},
ylabel near ticks,
yticklabel pos=right,
xmin = 1,
xmax = 9,
xshift = 1.5cm,
yshift = 0cm]
\addplot table [x = delay, y expr= \thisrow{sig}/0.7093, mark = none] {figures/Data_spectrum_temp.txt};
\node[] at (axis cs: 2,0.02) {(b)};
\end{axis}

\begin{axis}[
height = 5cm,
width = 7.5cm,
axis y line* = left,
xlabel={Frequency (THz)},
ylabel = {Intensity (normalized)},
xmin = 0.1,
xmax = 20,
ymin = 0,
ymax = 1.5,
yshift = -4.5cm]
\addplot table [x = nu, y expr= \thisrow{spec}/91.7733, mark = none] {figures/Data_spectrum_spec.txt};
\node[] at (axis cs: 2,1) {(c)};
\draw[black, dashed] (axis cs:4.514,0)--(axis cs:4.514,1.5);
\draw[black, dashed] (axis cs:5.114,0)--(axis cs:5.114,1.5);
\draw[black, dashed] (axis cs:6.080,0)--(axis cs:6.080,1.5);
\draw[black, dashed] (axis cs:6.824,0)--(axis cs:6.824,1.5);
\draw[black, dashed] (axis cs:7.616,0)--(axis cs:7.616,1.5);
\draw[black, dashed] (axis cs:8.354,0)--(axis cs:8.354,1.5);
\draw[black, dashed] (axis cs:9.086,0)--(axis cs:9.086,1.5);
\draw[black, dashed] (axis cs:9.83,0)--(axis cs:9.83,1.5);
\draw[black, dashed] (axis cs:10.51,0)--(axis cs:10.51,1.5);
\draw[black, dashed] (axis cs:11.25,0)--(axis cs:11.25,1.5);
\draw[black, dotted] (axis cs:18.15,0)--(axis cs:18.15,1.5);
\node[] at (axis cs: 17.5,0.3){Si};
\draw[<-, blue] (axis cs: 1,0.5) -- (axis cs: 3,0.5);
\end{axis}
\begin{axis}[
height = 5cm,
width = 7.5cm,
axis y line* = right,
axis x line = none,
ylabel = {$\alpha$ (cm$^{-1}$)},
xmin = 0.1,
xmax =20,
ymin = -6,
ytick = {-4,-3,-2,-1,0},
yticklabels = {4,3,2,1,0},
yshift = -4.5cm]
\addplot[color = red, mark = none] table [x = nu, y expr= -1*\thisrow{abs}] {figures/H2O_Hitran.txt};
\node[red] at (axis cs: 13,-1) {H$_2$O};
\draw[->, red] (axis cs: 17,0) -- (axis cs: 19,0);
\end{axis}
\end{tikzpicture}
\caption{(a) Temporal profile of the main THz pulse [identical to the pulse shown in Fig.~\ref{Fig:Profile}(a)]. (b) Trailing part of the THz pulse containing the spectroscopic features, with vertical scale adjusted for clarity.
(c) Reconstructed spectral intensity (blue curve) of the temporal profile of (a) and (b). The red curve represents the absorption coefficient $\alpha$ extracted from the HITRAN database for H$_2$O convoluted by a 0.1~THz FWHM Gaussian. The dashed lines serve as a guide to the eye by indicating most intense line positions. The dotted line indicates the position of a Si absorption band.}
\label{Fig:Spectrum}
\end{figure}

Let us now use the platform to investigate the spectroscopic features imprinted in the THz pulses through its beam path.
For these purpose, we increase the spectral resolution for a better identification of sharp spectral lines that may appear in the spectra.
To this end, we keep the temporal step of 2~fs in our acquisition (giving a spectral span of 250~THz), but we extend the temporal span to 10~ps, thus reducing the spectral resolution down to 0.1~THz.
The acquisition time using these parameters and averaging over 10 shots for each data point is below 10~minutes.
Resulting temporal and spectral profiles (still without any sample at the THz focus) are presented in Fig.~\ref{Fig:Spectrum}.
Figure~\ref{Fig:Spectrum} (a) and (b) show different parts of the same signal with adjusted vertical scales.
It can be seen that after the main pulse (a) [identical to the pulse shown in Fig.~\ref{Fig:Profile}(a)], a weak signal is present in the trailing part of the pulse (b). 
In our setup, this signal is mainly attributed to the presence of ambient water vapor~\cite{Babilotte2016} in the dry air enclosure that is kept at 9\% relative humidity.
In the spectral domain, it corresponds to absorption bands in the spectrum, as can be seen in Fig.~\ref{Fig:Spectrum}(c).
Even if the amplitude of this signal is significantly smaller than the main peak, it carries the essential spectral information on the water absorption bands.
This amplitude ratio highlights the need for a high signal to noise ratio, which has been evaluated to about 100 for our acquisition system.
Figure~\ref{Fig:Spectrum}(c) compares the position of the measured H$_2$O absorption bands with the absorption coefficient $\alpha$ extracted from the HITRAN database\cite{GORDON2021}, convoluted with a 0.1~THz FWHM Gaussian to match our spectral resolution.
The position of the main lines are in excellent agreement, giving a good standard to validate our measurements.
Another weak absorption band in the measured spectrum, close to 18.1~THz, can be attributed to a phonon resonance in the silicon wafer\cite{Kaltenecker2019}.

\subsection{THz time-domain spectroscopy of polystyrene}
\begin{figure}[!tbp]
\tikzsetnextfilename{Sample_spectrum}
\begin{tikzpicture}
\begin{axis}[
height = 4cm,
width = 8.5cm,
xlabel={Delay (ps)},
ylabel = {Signal (normalized)},
xmin = -0.5,
xmax = 1]
\addplot[red] table [x = delay, y expr= \thisrow{sig}/0.7093, mark = none] {figures/Data_spectrum_temp.txt};
\addplot[blue] table [x = delay, y expr= \thisrow{sig}/0.7093, mark = none] {figures/Data_sample_temp.txt};
\node[] at (axis cs: -0.4,0.9) {(a)};
\end{axis}
\begin{axis}[
height = 4cm,
width = 8.5cm,
xlabel={Frequency (THz)},
ylabel = {Intensity (normalized)},
xmin = 0.1,
xmax = 40,
ymin = 0,
ymax = 1.1,
yshift = -3.5cm]
\addplot table [x = nu, y expr= \thisrow{spec}/87.6353, mark = none] {figures/Data_sample_spec.txt};
\node[] at (axis cs: 2.5,0.9) {(b)};
\end{axis}
\begin{axis}[
height = 4cm,
width = 7.6cm,
ylabel = {Transmission},
xlabel={Frequency (THz)},
axis y line* = left,
xmin = 10,
xmax = 40,
ymin = 0,
ymax = 1,
yshift = -7cm]
\addplot[blue] table [x = nu, y expr= \thisrow{trans}, mark = none] {figures/Data_sample_trans.txt};
\node at (axis cs: 12.5,0.6) {(c)};
\draw[black, dashed] (axis cs:16.21,0)--(axis cs:16.21,1.5);
\draw[black, dashed] (axis cs:21.01,0)--(axis cs:21.01,1.5);
\draw[black, dashed] (axis cs:22.72,0)--(axis cs:22.72,1.5);
\draw[black, dashed] (axis cs:27.21,0)--(axis cs:27.21,1.5);
\draw[<-, blue] (axis cs: 11,0.15) -- (axis cs: 15,0.15);
\draw[->, red] (axis cs: 35,0.8) -- (axis cs: 39,0.8);
\end{axis}
\begin{axis}[
height = 4cm,
width = 7.6cm,
axis y line* = right,
axis x line = none,
ylabel = {Absorption (arb. u.)},
ytick = {-4,-3,-2,-1,0},
yticklabels = {4,3,2,1,0},
xmin = 10,
xmax = 40,
ymin = -4,
ymax = 0,
yshift = -7cm]
\addplot[red] table [x = nu, y expr= -\thisrow{abs}, mark = none] {figures/Data_sample_ftir.txt};
\node at (axis cs: 12.5,0.6) {(c)};
\end{axis}
\end{tikzpicture}
\caption{(a) Temporal profile of the THz electric field with (blue) and without (red) polystyrene sample both normalized to the reference spectrum peak value. (b) Retrieved intensity spectrum with the polystyrene sample. (c) Reconstructed transmission spectrum with polystyrene sample (blue) and polystyrene absorption spectrum extracted from Bruker ATR-Polymer Library (red). The dashed lines serve as a guide to the eye and indicate the positions of the main absorption lines.}
\label{Fig:sample_Spectrum}
\end{figure}
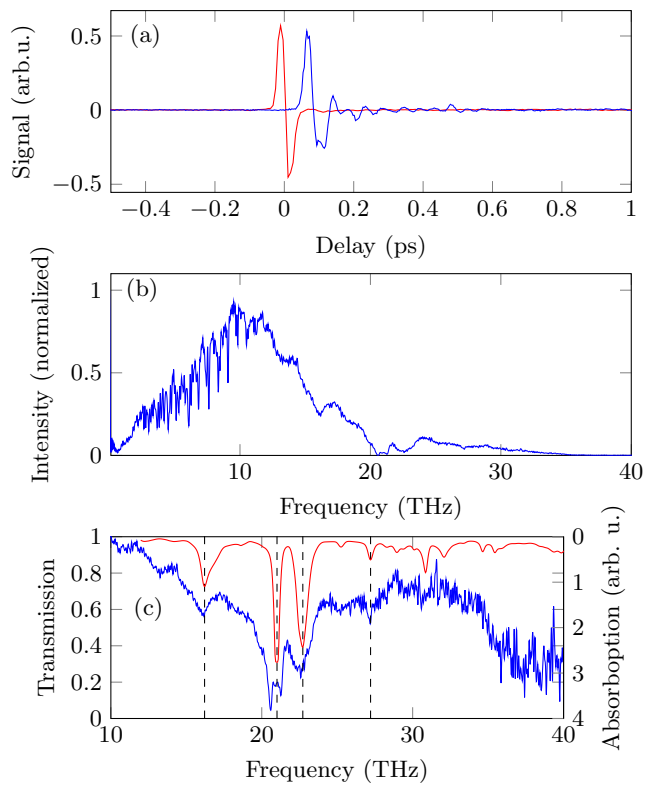

We finally perform time-domain spectroscopy of a polystyrene sheet and compare it with data extracted from a Fourier Transform Infra- Red (FTIR) spectroscopy database.
Such sheets are frequently used as a reference to calibrate FTIR spectrometers in both line positions and amplitudes.
The acquisitions parameters are the same as in the previous section, that is, the spectral resolution is set to 0.1~THz.
Figure~\ref{Fig:sample_Spectrum}(a) shows the measured temporal profiles.
The signal with sample is delayed due to global change of the group velocity inside the polystyrene sheet.
The polystyrene dispersion stretches the signal significantly, leading to some oscillations appearing between the main pulse and about 0.6~ps.
Absorption leads to the appearance of spectral bands between 10~THz and 30~THz, as well as the water lines discussed previously.

To obtain a better visualization of the polystyrene spectral bands we calculated the transmission spectrum of the sample by dividing the sample spectrum by the reference spectrum.
The obtained transmission spectrum is shown in Fig.~\ref{Fig:sample_Spectrum}(c), and compared with a reference spectrum from the OPUS software Bruker ATR-Polymer-Library\cite{opus}.
The main line positions between 15~THz and 30~THz are in good agreement. For lower frequencies there are no absorption lines to compare, and for higher frequencies the signal-to-noise ratio becomes too low to perform any comparison.
The signal near the 21~THz line drops to the noise level [see Fig. \ref{Fig:sample_Spectrum}(b)], thus the transmission value calculated around this frequency might not be relevant. It may explain why the measured line shape does not match the one observed in the FTIR spectrum.

\section{Conclusion}
We have presented an air photonics based terahertz platform with micro-controller interface for basic operations and data acquisition.
The platform is thus easy to use and customizable for future experiments.
Our detection setup does not require a lock-in amplifier, but only relies on a peak detector and the application of a high voltage bias of switchable sign.
Using this simple setup we report a signal to noise ratio above 100. 
The platform produces short THz pulses with a broadband spectrum reaching 30~THz and a peak electric field strength in the MV/cm range.
This makes for a versatile platform usable for THz-TDS but also for other type of applications employing THz pulses as pump or probe such as THz streaking\cite{ArdanaLamas2016}, molecular orientation\cite{Xu2020} or nonlinear spectroscopy\cite{Tcypkin2021}.

Comparison with comprehensive numerical simulations of both the THz source and detection reveals the origin of the characteristic spectral shape obtained with ABCD measurements.
We also demonstrated the importance of taking into account pump propagation effects before the plasma for THz generation is formed, in particular in the BBO crystal and waveplate. Such detailed confrontation of theoretical results and measurements performed on a given setup is unprecedented.
We thus expect strong synergistic effects between experiments and numerical simulations in the future improvement and exploitation of our THz platform.

\begin{acknowledgments}
We thank B. Moge, I. Aguili, F. Khalid Balyos and C. Clavier for technical support.
We thank P. U. Jepsen and  B. Zhou from the Technical University of Danemark for fruitful discussions.
We acknowledge financial support from CNRS, the ANR ALTESSE2 (ANR-19-ASMA-0007) project, and the Qatar National Research Fund (NPRP 12S-0205-190047). Numerical simulations were performed using resources at Grand Équipement National De Calcul Intensif
(GENCI) (A0100507594).
\end{acknowledgments}

\section*{Author declarations}
\subsection*{Conflict of Interest}
The authors have no conflicts to disclose.

\section*{Data Availability Statement}
The data that support the findings of this study are available from the corresponding author upon reasonable request.



%
%

%


\bibliography{biblio}

\end{document}